\documentclass[11pt]{article}
\usepackage[top=3cm, bottom=3.9cm, left=3cm, right=3cm]{geometry}
\usepackage{graphicx}
\usepackage{calc}
\usepackage{amsmath}
\usepackage{bbm}
\usepackage{mathtools, cuted}
\usepackage{multicol}
\usepackage{amsfonts}
\usepackage{amsthm}
\usepackage{amssymb}
\usepackage{amsbsy}
\usepackage[T1]{fontenc}
\usepackage{lipsum}
\usepackage{authblk}
\usepackage{textcomp}
\usepackage{subfigure}
\usepackage{epstopdf}
\usepackage{lmodern}
\usepackage[overload]{empheq}
\usepackage{framed}
\usepackage[subfigure]{tocloft}
\usepackage{subeqnarray}
\usepackage{alphalph}
\usepackage[refpage,cfg,prefix]{nomencl}
\usepackage{multirow}
\usepackage{xifthen}
\usepackage{enumerate}
\usepackage[bottom]{footmisc}
\usepackage{natbib}  
\usepackage{wasysym}
\usepackage{bibentry}
\usepackage{verbatim}
\usepackage{url}
\usepackage{setspace}

\usepackage{lipsum}                     
\usepackage{xargs}                      
\usepackage[pdftex,dvipsnames]{xcolor}  
\usepackage[colorinlistoftodos,prependcaption,textsize=tiny]{todonotes}
\newcommandx{\todoadd}[2][1=]{\todo[linecolor=blue,backgroundcolor=blue!25,bordercolor=blue,#1]{#2}}
\newcommandx{\todoinf}[2][1=]{\todo[linecolor=yellow,backgroundcolor=yellow!25,bordercolor=OliveGreen,#1]{#2}}
\newcommandx{\seen}{\todo[linecolor=OliveGreen,backgroundcolor=OliveGreen!25,bordercolor=OliveGreen]{Review already checked}}
\newcommandx{\notseen}{\todo[linecolor=red,backgroundcolor=red!25,bordercolor=red]{Review NOT checked}}
\newcommandx{\todoimp}[2][1=]{\todo[linecolor=Plum,backgroundcolor=Plum!25,bordercolor=Plum,#1]{#2}}
\newcommandx{\todohid}[2][1=]{\todo[disable,#1]{#2}}

\usepackage{color,colortbl}
\usepackage{relsize}

\makenomenclature

\newcommand{\D}[2]{\frac{\partial #1}{\partial #2}}
\newcommand{\DD}[2]{\frac{\partial^2 #1}{\partial #2^2}}

\newcommand{\Dstraight}[2]{\frac{\ud#1}{\ud#2}}
\newcommand{\ud}{\mbox{d}}

\newcommand{\R}{\ensuremath{\mathbb{R}}}

\newcommand{\LR}[1]{\left(#1\right)}
\newcommand{\LRb}[1]{\left\lbrace#1\right\rbrace}

\newcommand{\LRs}[1]{\left [#1\right ]}
\newcommand{\st}{\ensuremath \, | \,}

\newcommand{\IM}[1]{\ensuremath \text{Im}\LRs{#1}}
\newcommand{\RE}[1]{\ensuremath \text{Re}\LRs{#1}}
\newcommand{\maxeig}{\rho(\underbar{$\mu$})}
\newcommand{\lagra}{\mathcal{L}}

\newif\ifletter

\begin{document}

\title{\textbf{The invasion speed of cell migration models with realistic cell cycle time distributions}}

	
	\author[1]{Enrico Gavagnin\footnote{Corresponding author: e.gavagnin@bath.ac.uk}}
	\author[2]{Matthew J. Ford}
	\author[3]{Richard L. Mort}
	\author[1]{Tim Rogers}
	\author[1]{Christian A. Yates} 
	\affil[1]{\small{\textit{Department of Mathematical Sciences}}\\ \small{\textit{University of Bath, Claverton Down, Bath, BA2 7AY, UK}}}
	\affil[2]{\small{\textit{Centre for Research in Reproduction and Development}}\\ \small{\textit{McGill University, Montr\'eal, H3G 1Y6, Qu\'ebec}}}
	\affil[3]{\small{\textit{Division of Biomedical and Life Sciences}}\\ \small{\textit{Faculty of Health and Medicine}}\\ \small{\textit{Lancaster University, Bailrigg, Lancaster LA1 4YG, UK}}}

\date{}
\maketitle
\vspace{-25pt}
\begin{abstract}
Cell proliferation is typically incorporated into stochastic mathematical models of cell migration by assuming that cell divisions occur after an exponentially distributed waiting time. Experimental observations, however, show that this assumption is often far from the real cell cycle time distribution (CCTD). Recent studies have suggested an alternative approach to modelling cell proliferation based on a multi-stage representation of the CCTD. 

In order to validate and parametrise these models, it is important to connect them to experimentally measurable quantities. In this paper we investigate the connection between the CCTD and the speed of the collective invasion. We first state a result for a general CCTD, which allows the computation of the invasion speed using the Laplace transform of the CCTD. We use this to deduce the range of speeds for the general case. We then focus on the more realistic case of multi-stage models, using both a stochastic agent-based model and a set of reaction-diffusion equations for the cells' average density. By studying the corresponding travelling wave solutions, we obtain an analytical expression for the speed of invasion for a general  $N$-stage model with identical transition rates, in which case the resulting cell cycle times are Erlang distributed. We show that, for a general $N$-stage model, the Erlang distribution and the exponential distribution lead to the minimum and maximum invasion speed, respectively. This result allows us to determine the range of possible invasion speeds in terms of the average proliferation time for any multi-stage model.

\end{abstract}
\section{Introduction}

Cellular invasion is a process of fundamental importance in numerous morphogenetic and pathological mechanisms. Important examples include embryonic development \citep{gilbert2003med,keller2005cmd}, wound healing \citep{maini2004twm,deng2006rma} and tumour invasion \citep{hanahan2000hoc}. 

During an invasion, cells' behaviour can be characterised through a variety of different biological mechanisms, such as chemotaxis \citep{keynes1992rca,ward2003dbd}, cell-cell adhesion \citep{niessen2007tja, trepat2009pfd} and cell-cell attraction \citep{yamanaka2014vas}. Moreover, there is evidence that cells behave differently within the wave of invasion. For example, in cell migration neural of crest cells in the developing embryo, a small group of cells at the front of the wave, called \textit{leaders}, are responsible for the exploration of the environment, while the remaining cells, called \textit{followers}, simply undergo an adhesive behaviour \citep{mclennan2012mmc,mclennan2015ncm,schumacher2017shc}.  

Understanding how the properties of the individual cells contribute to the formation and the propagation of the wave is fundamental. In fact, this can reveal the micro-scale mechanisms that are responsible for a given phenomenological aspect, and hence suggest effective therapeutic approaches to inhibit, or enhance, cell migration by interrupting the cell cycle \citep{sadeghi1998vem, gray2007mbn,haass2017cct}. 

Despite the large variety of actions and interactions which cells can undergo, there are at least two aspects of cells' behaviour which are essential in order for the invasion to take place. These are cell dispersal and cell proliferation \citep{simpson2007sic,mort2016rdm}. If one of these two aspects does not occur properly, the impact on the collective invasion is typically evident and it can affect the success of the colonisation. For example, \citet{mort2016rdm} show using an experimental and a modelling approach that the lack of  colonisation of mouse melanoblasts is probably driven by reduced proliferation.

Extensive research has focused on the effect that cell dispersal and proliferation behaviours have on the speed of the invasion, $c$. The common approach makes use of simple mathematical models which typically take the form of a stochastic agent-based model (ABM) \citep{anderson1998cdm,deutsch2007cam} or a deterministic partial differential equation (PDE) \citep{murray2007mbi,wise2008tdm}. By computing the invasion speed of the model, either analytically or numerically, it is possible to link the parameters which modulate the diffusivity and proliferation with the speed of invasion.

Many studies have investigated this link in more general contexts, beginning with the work of \citet{fisher1937waa} on the spread of a favoured gene through a population and including more recent studies in ecology \citep{holmes1994pde,elliott2012dps}. From these studies, it is well known that, when dispersion and proliferation occur with rates $\alpha$ and $\lambda$, respectively,  we have that the invasion speed is proportional to the square root of the product of the rates, i.e. $c\propto \sqrt{\alpha \lambda}$ \citep{fisher1937waa}. 

\begin{figure*}[h!!]
 \begin{tabular}{m{0.5\hsize}m{0.5\hsize}}
\subfigure[][]{\includegraphics[width=0.4 \columnwidth]{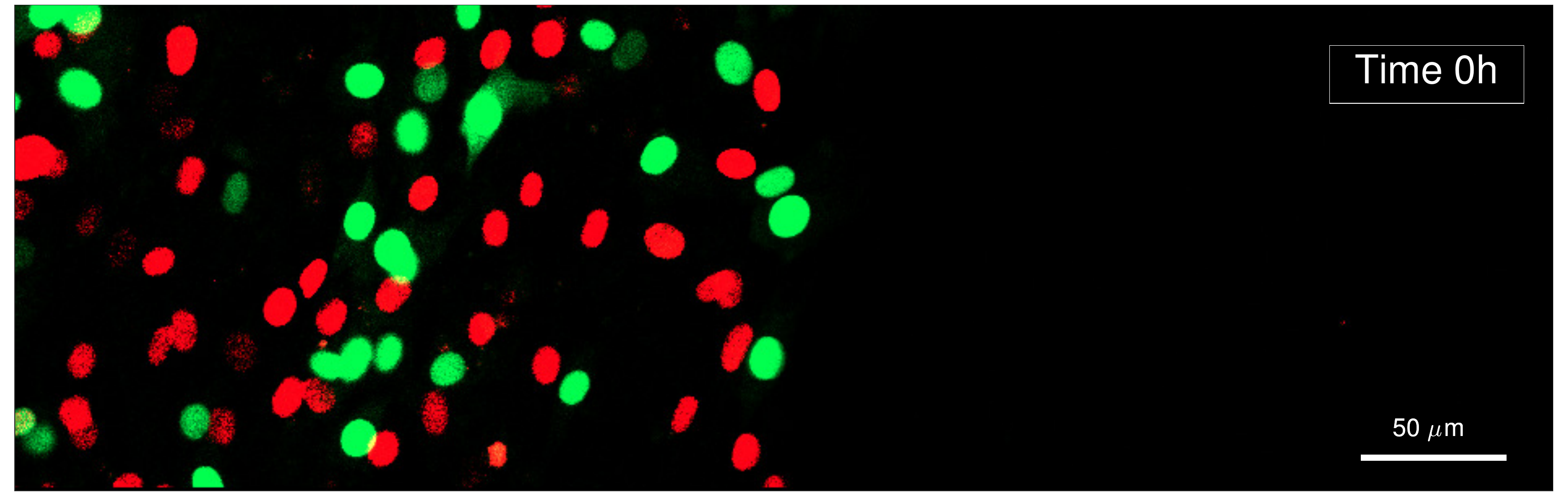} }\\[27pt]
\subfigure[][]{\includegraphics[width=0.4 \columnwidth]{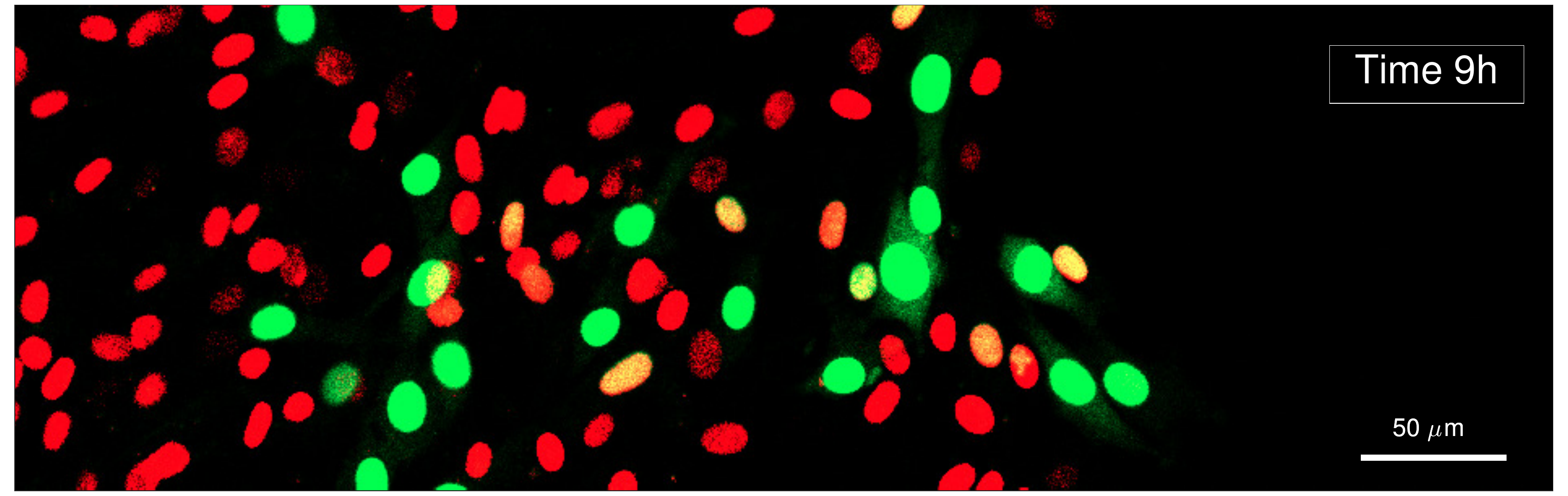} }\\[27pt]
\subfigure[][]{\includegraphics[width=0.4 \columnwidth]{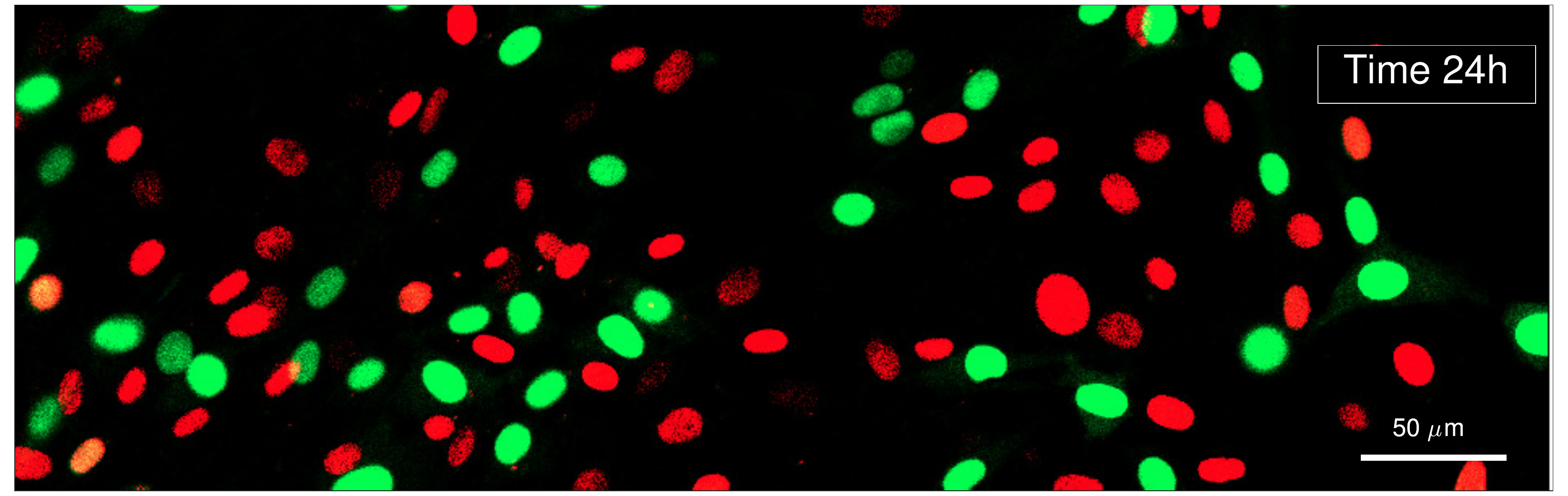}}\\[27pt]
\end{tabular}
\begin{tabular}{m{0.5\hsize}m{0.5\hsize}}
\subfigure[][]{\includegraphics[width=0.4 \columnwidth]{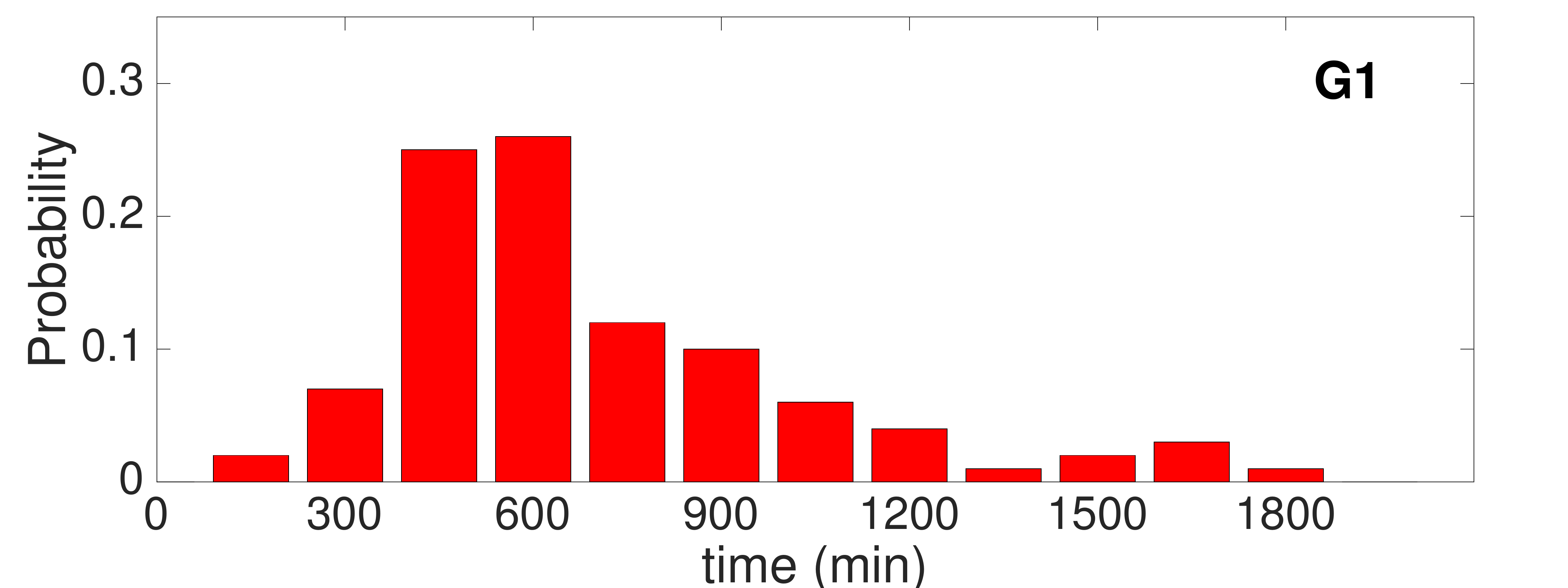} }\\
\subfigure[][]{\includegraphics[width=0.4 \columnwidth]{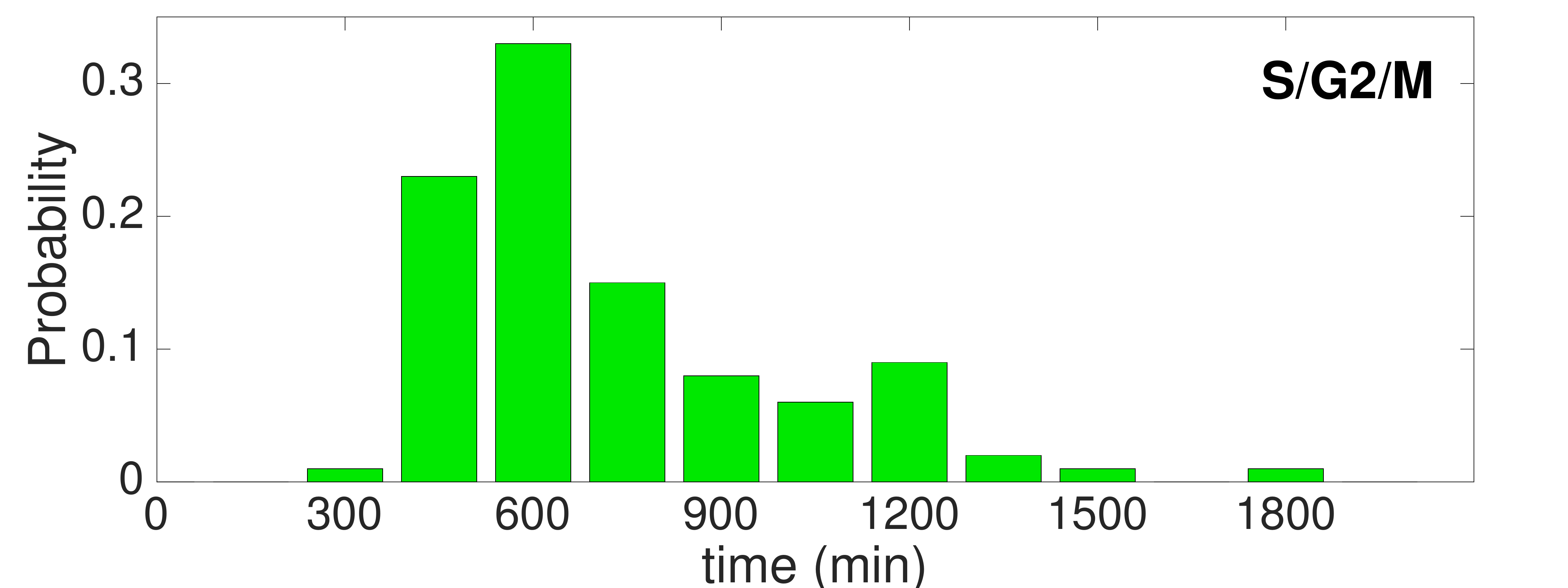} }\\
\subfigure[][]{\includegraphics[width=0.4 \columnwidth]{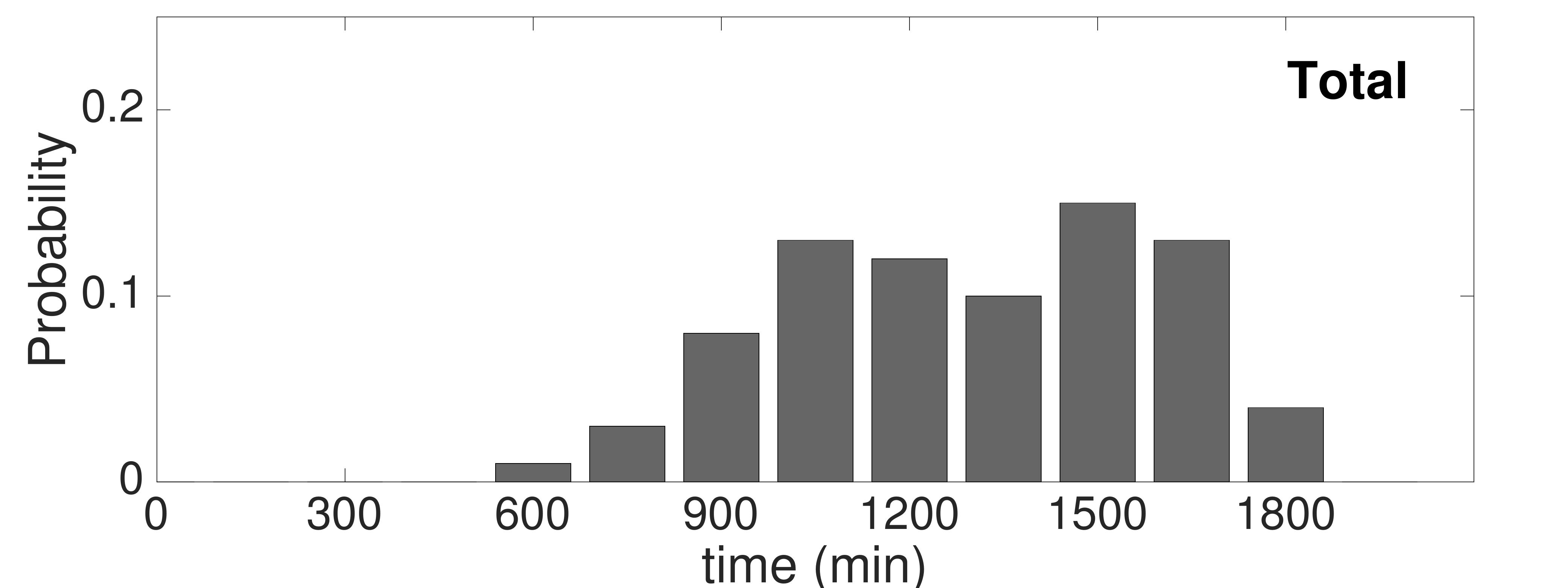}}
\end{tabular}
\caption{Panels (a-c): Mouse NIH-3T3 fibroblasts with Fucci2a status migrating into open space \citep{mort2014fbc}. The Fucci2a system incorporates genetically encoded probes that highlight in red the nuclei of cells in the G1 phase and in green those of cells in one of the other phases, S/G2/M. Panels (d-f): experimental distributions of the time length of the G1 phase (panel (d)), S/G2/M phases (panel (e)) and total CCTD (panel (f)). Both the G1 and S/G2/M distributions  show a clear non-monotonic trend, which indicates that are not exponentially distributed. To capture both these non-monotonicities using a MSM for the CCTD, a minimum of four stages is required, two for each of the two phases.}
\label{fig:evidence} 
\end{figure*}

It is important to notice the that great majority of the literature on the speed of invasion of travelling waves is based on the assumption that proliferation events occur as independent Poisson processes \citep{simpson2007sic,mort2016rdm}. In the context of cell migration, this is equivalent of assuming that cells proliferate after an exponentially distributed random time. However, experimental observations show that the cell cycle time distribution (CCTD) is typically non-monotonic and it differs substantially from an exponential distribution (see Figure \ref{fig:evidence}\,(f) for an example) \citep{golubev2016aie,yates2017msr,chao2018ecc}. 

There is a vast literature regarding the appropriate represenation of  the CCTD \citep{csikasz2006agm,gerard2009tso,powathil2012mec}. One class of represenations, known as multi-stage models (MSMs), have gained particular attention in several recent studies \citep{golubev2016aie,yates2017msr,vittadello2018mmc,chao2018ecc}. The main idea of MSMs is to partition the cell cycle into $N$ sequential stages. As time evolves, each cell can transit from one stage, $i$, to the next one, $i+1$, after an exponentially distributed waiting time with parameter $\lambda_i$. When a cell is found at the last stage, $N$, it can proliferate with rate $\lambda_N$, which leads the cell to split into two daughter cells, both initialised at the first stage. The main motivation that makes MSMs mathematically appealing is the Markov property of the exponentials which simplifies both the analytical investigation of the model and its computational implementation. Moreover, MSMs lead to a CCTD called hypoexponential distribution which has been shown to provide an excellent agreement with experimental data \citep{golubev2016aie,yates2017msr,chao2018ecc}.

It is important to think about the multi-stage partition merely as a mathematical tool, rather than a biologically realistic representation.  In particular, the \textit{stages} of the MSMs should not be confused with the biological \textit{phases} of the cell cycle which, in general, are not exponentially distributed (see Figure \ref{fig:evidence}) \citep{chao2018ecc}.

Whilst previous studies have investigated MSMs extensively in the case of spatially uniform scenarios \citep{yates2017msr}, there is still little understanding about the effect which MSMs have on invading waves of cells. In particular, it is not clear how, and to what extent, a multi-stage representation of the CCTD can impact on the speed of invasion. 

The most recent progress on this was made by \citet{vittadello2018mmc}. In their work, the authors derive an analytical expression for the invasion speed of a 2-stage MSM in terms of the two rates of stage transition, $\lambda_1,\lambda_2$, and the diffusion coefficient of cells, $D$:
\begin{equation}
	\label{eq:vittadello}
	c=\sqrt{2 D \LRs{-\lambda_1 -\lambda_2 +\sqrt{\lambda_1^2+6\lambda_1 \lambda_2 +\lambda_2^2}\,}} \, .
\end{equation}
The findings of \citet{vittadello2018mmc} provide useful insights in the qualitative effect of the MSMs. However a general expression for the invasion speed, as in equation \eqref{eq:vittadello}, but for biologically realistic MSMs, which typically have ten or more stages \citep{yates2017msr,chao2018ecc}, is not feasible analytically. Hence, there are important questions about the quantitative effect of MSMs on the invasion speed which remain unanswered. In particular, the range of variability in speed for a general $N$-stage MSM has yet to be studied.

To investigate the effect of incorporating a general CCTD into the invasion models, we follow two distinct approaches. Firstly, we formulate a generalisation of the Fisher-KPP equation which describes the cell population as age-structured. By studying the traveling wave solutions of the model, we derive an implicit equation for the speed of invasion in terms of the Laplace transform of the CCTD. This allows us to show that the speed for a general proliferation time distribution can be arbitrary large, whereas we obtain an expression for the minimum possible speed.

In the second part of the paper, we focus our attention on MSMs. We study a spatially extended ABM which is designed to mimic cell invasion on a regular two-dimensional lattice. For each agent, we implement a general $N$-stage MSM to simulate the stochastic waiting time before the agent attempts to divide into two daughters. Through a mean-field closure approximation on the average agent density, we derive a system of $N$ reaction-diffusion PDEs which represents a generalisation of the model of \citet{vittadello2018mmc}. By applying the front propagation method \citep{vansaarlos2003fpu} to the system of PDEs, we reduce the computation of the invasion speed to an eigenvalue problem in terms of the rates of transition between consecutive stages, $\lambda_i$. We use this result to study the case of identical transition rates, that corresponds to modelling the CCTD as Erlang.  In this case we provide the exact analytical expression for the speed. By combining our findings, we formulate a result for the maximum and minimum speed for a general $N$-stage MSM.

The paper is organised as follows. In Section \ref{sec:age_struc} we define the age-structured model and we derive the implicit equation for the invasion speed for general CCTD. In Section \ref{sec:models} we define two MSMs: a stochastic ABM and the corresponding mean-field approximation. In Section  \ref{sec:analysis} we explain how to apply the front propagation method and we state the eigenvalue problem. We present our results on  Erlang distributed cell cycle times and the general hypoexponential case in Section \ref{sec:results}. We conclude in Section \ref{sec:conclusions} with a brief discussion of this work and future challenges.

\section{Age-Structured Model}
\label{sec:age_struc}
The Fisher-KPP equation implicitly assumes Markov dynamics for the individual cells making up the population, implying a cell cycle time with an exponential distribution \citep{fisher1937waa}. One way to adapt the model to allow for an arbitrary cell cycle time distribution is through the addition of age-structure. Cells have an associated age, denoted by $a$, which takes values in the positive real numbers and increases as time evolves. Cells diffuse, with diffusivity $D$, and proliferate with an age-dependent rate, $h(a)$.  

We can write down a simple linear PDE for the density of cells with age $a$ and spatial location $x$ at time $t$, $C(a,x,t)$, as follows 
\begin{equation}
\begin{split}
\frac{\partial}{\partial t} C(a,x,t) &= -\frac{\partial}{\partial a} C(a,x,t) + D\frac{\partial^2}{\partial x^2} C(a,x,t) - h(a)C(a,x,t)\\
C(0,x,t)&=2\int_0^\infty h(a)C(a,x,t)\textrm{d}a\,.
\end{split}
\label{asm}
\end{equation}
The function $h(s)$ is the hazard rate, related to the probability density function $f(s)$ of the age at which cells divide (i.e. the CCTD) via 
\begin{equation}
h(s)=\frac{f(s)}{\int_s^\infty f(a)\,\textrm{d}a}\,,\qquad f(s)=h(s)\exp\left(-\int_0^s h(a)\textrm{d}a\,\right)\,.
\label{hf}
\end{equation}
The boundary condition for $C(0,x,t)$ gives the density of newborn cells as twice the total rate of cell division. Note that we have neglected from (\ref{asm}) any non-linear terms arising from crowding effects, as these are not relevant to the speed of the front propagation. This model is a simple spatial adaptation of the McKendrick-Von Foerster equation for growing age-structured populations, and has been studied before \citep{webb1985tna, al2002mtf,gabriel2012cas}. 

In our first result, we show that the speed of propagation for the model \eqref{asm} is determined by the Laplace transform of the CCTD, defined by
\begin{equation}
\mathcal{L}\{f\}(s)=\int_0^\infty e^{-sa}f(a)\,\textrm{d}a\,.
\end{equation}

\paragraph{Theorem 1}
If $\lim_{s\to\infty}\mathcal{L}\{f\}(s)<1/2$ then the PDE (\ref{asm}) admits travelling wave solutions with propagation speed $c>2\sqrt{D\lambda}$, where $\lambda>0$ is the unique solution to 
\begin{equation}
\label{eq:theo0}
\mathcal{L}\{f\}(\lambda)=1/2\,.
\end{equation}

\begin{proof}
The system (\ref{asm}) is seperable, hence we seek solutions of the form $C(a,x,t)=v(a)w(x-ct)$, corresponding to a travelling wave with speed $c$ and internal age structure given by $v$. Inserting into (\ref{asm}) and rearranging, we find
\begin{equation}
c\frac{w'}{w}+D\frac{w''}{w}=\frac{v'}{v}+h\,.
\end{equation}
The left-hand side here is a function only of $x-ct$, whilst the right-hand side is a function only of $a$. We thus determine that both are equal to a constant, say $-\lambda$. The $w$ equation becomes
\begin{equation}
\lambda w + cw'+Dw'' = 0\,,
\end{equation}
which is well-known as the linear part of the Fisher-KPP equation, admitting travelling wave solutions for all $c>2\sqrt{D\lambda}\,.$ The equation for $v$ has solution
\begin{equation}
v(a)=v(0)\exp\left(-a\lambda-\int_0^ah(\alpha)\textrm{d}\alpha\right)\,.
\end{equation}
The boundary condition then gives us 
\begin{equation}
1=2\int_0^\infty h(a) \exp\left(-a\lambda-\int_0^ah(\alpha)\textrm{d}\alpha\right)\textrm{d}a\,,
\end{equation}
from which the definition of the hazard rate, equation \eqref{hf}, identifies the result $1=2\mathcal{L}\{f\}(\lambda)$. Uniqueness of the solution (when one exists) follows from the monotonicity of the Laplace transform of a probability density.
\end{proof}

We can use the previous result to investigate the range of speeds for an arbitrary CCTD with a given mean, $\bar{\mu}$. By using Jensen's inequality we have that for any positive supported $f$ with mean $\bar{\mu}$
\begin{equation}
	\mathcal{L}\{f\}(\lambda)\le e^{-\lambda \bar{\mu}}=\mathcal{L}\{\delta_{\bar{\mu}}\}(\lambda) \, ,
\end{equation} 
where $\delta_{\bar{\mu}}$ is the Dirac delta function concentrated at $\bar{\mu}>0$. From the monotonicity of the Laplace transform of a probability density, it follows that the minimum speed is obtained by using $f=\delta_{\bar{\mu}}$, which gives
\begin{equation}
	\label{eq:speed_min}
	c\ge 2 \sqrt{\frac{D \ln 2}{\bar{\mu}}} \, 
\end{equation} 

We now use Theorem 1 to show that there is no upper bound for the speed of invasion of a general CCTD with a given mean. Consider the set of probability density functions defined as
\begin{equation}
	f_\varepsilon (x)=\frac{1}{2} \LR{\delta_{\varepsilon \bar{\mu}}+\delta_{(2-\varepsilon) \bar{\mu}}}\, ,
\end{equation}
where $\varepsilon\le 1$. It is immediate to observe that each member of this set of functions have mean $\bar{\mu}$ and Laplace transform given by:
\begin{equation}
\label{eq:laplace_max}
	\mathcal{L}\{f_\varepsilon\}(\lambda)=\frac{1}{2} \LR{e^{-\lambda\varepsilon \bar{\mu}}+e^{-\lambda(2-\varepsilon) \bar{\mu}}} \, .
\end{equation} 
By substituting the expression \eqref{eq:laplace_max} into equation \eqref{eq:theo0} and rearranging, we obtain the implicit equation for $\lambda$ given by
\begin{equation}
\label{eq:impl_eq_max}
	\lambda \varepsilon \bar{\mu}=-\ln \LR{1-e^{-2\lambda \bar{\mu}}} \, .
\end{equation}
The right-hand side of equation \eqref{eq:impl_eq_max} is a strictly decreasing function of $\lambda$ that converges to $0$ as $\lambda\rightarrow \infty$. Therefore, we can always choose $\varepsilon$ small enough so that the solution of equation \eqref{eq:impl_eq_max} is arbitrarily large. 

This demonstrates that, assuming that the CCTD is a general function with mean $\bar{\mu}$ and positive support, the range of possible invasion speeds is given by
\begin{equation}
\label{eq:range_general}
	c\in \left[ 2 \sqrt{\frac{D \ln 2}{\bar{\mu}}} , \infty \right ) \, .
\end{equation}

The result in Theorem 1 is important because it establishes the connection between a general CCTD and the corresponding invasion speed. However, for some particular classes of distributions, solving equation \eqref{eq:theo0} analytically can be challenging and the method of this Section does not provide any deeper insights. In particular, this is true for hypoexponential distributions, which are of special interest in the context of cell proliferation. In the next three sections we further explore this class of distributions by adopting an \textit{ad hoc} modelling approach.

\section{Multi-Stage Models}
\label{sec:models}
In this section we introduce the two MSM that we will use throughout the paper. Firstly, we define a discrete ABM, in which the multi-stage representation of the CCTD is implemented as a stochastic feature of each cell at the microscale. Secondly, we introduce a system of deterministic PDEs describing the average cell density in a macroscopic manner.

\paragraph{The ABM}
We consider a continuous-time ABM on a two-dimensional regular square lattice, with a given spacing denoted by $\Delta$. Each cell is modelled as a single agent which moves and proliferates. Volume exclusion is incorporated by allowing at most one agent to occupy a given lattice site.

Agents move according to a simple excluding random walk on the lattice. Each agent attempts a movement after an exponentially distributed waiting time with rate $\alpha$. When this happens, a new position is chosen uniformly from one of the four nearest neighbouring sites and the movement takes place only if the selected site is empty. The  event is aborted otherwise. 

We implement cell proliferation using a MSM. We divide the cell cycle into $N$ sequential stages. Agents at one of the first $N-1$ stages, $i=1,\dots, N-1$, move to the next stage after an exponentially distributed waiting time of rate $\lambda_i$. Agents at the last stage, $N$, can attempt a proliferation event, after a further exponentially distributed waiting time of rate $\lambda_N$. In order to attempt a proliferation event, a target site is selected uniformly at random from one of the four nearest neighbouring sites. If such site is empty, a new first-stage agent is located on it, and the proliferating agent is returned to the first stage. If the target site is occupied, the proliferation event is aborted and the proliferating agent remains at the last stage\footnote{Alternatively, we could choose to return the proliferating agent to the first stage every time an abortion occurs. This model has been studied in \citet{yates2017msr} for homogeneously distributed agents. This modification does not substantially change our results. This is because our analysis of the speed of the wave front is based on low density regions, where abortion of events do not play an important role. For this reason, we decided to focus only on the stated version of the model.}. 

We simulate the cell invasion by populating the lattice with first-stage agents located at random in the first 10 columns on the left of the domain. We impose zero flux boundary conditions on the $x$-direction and periodic boundary conditions on the $y$-direction. Agents are displaced uniformly at random in the vertical direction, so  we can reduce the dimensionality of the problem by considering the average column density \citep{simpson2009mss}. 

\paragraph{The PDE model} 
Here we define the continuous model for the average column density which will be the object of the invasion speed analysis.

We denote with $S_i(x,t)$ density of $i$-stage agents in the column $x$ at time $t$, averaged over multiple realisations fo the ABM. Let $C(x,t)$ be the total density of column $x$ at time $t$, i.e. 
\begin{equation}
	C(x,t)=\sum_{i=1}^N S_i(x,t) \, .
\end{equation}

By writing down the master equation of $S_i$, for $i=1, \dots , N$ and taking the limit as $\Delta\rightarrow 0$, while keeping $\alpha \Delta^{2}$ constant, one can derive a system of reaction-diffusion PDEs for the column densities of the different stages:
\begin{equation}
\label{eq:PDE}
	\begin{cases}
		\D{S_1}{t}&= D \D{}{x}\LRs{(1-C)\D{S_1}{x}+S_1 \D{C}{x}}+2\lambda_N (1-C) S_N - \lambda_1 S_1 \\[4 pt]
		\D{S_i}{t}&= D \D{}{x}\LRs{(1-C)\D{S_i}{x}+S_i \D{C}{x}}+\lambda_{i-1} S_{i-1} - \lambda_i S_i  \qquad \quad  \text{for $i=2, \dots, N-1$}\\[4 pt]
		\D{S_N}{t}&= D \D{}{x}\LRs{(1-C)\D{S_N}{x}+S_N \D{C}{x}}+ \lambda_{N-1} S_{N-1} -\lambda_N (1-C) S_N \, ,
	\end{cases}
\end{equation}
where $D= \lim_{\Delta \rightarrow 0} \frac{\alpha \Delta^2 }{4}$. A detailed derivation for the three-stage model on an hexagonal lattice can be found in \citet{simpson2018smc}.

System \eqref{eq:PDE} consists of a set of reaction-diffusion PDEs with non-linearities in both the diffusion and the proliferation terms due to the introduction of volume exclusion. Specifically, the term $(1-C)$ accounts for the reduction in rate due to volume exclusion. Notice that by summing all the equations in \eqref{eq:PDE}, we recover simple diffusion for the total agent density. However, it is not possible to obtain a closed PDE for the total agent density without further assumptions \footnote{This can be done, for example, by assuming that the system has reached an equilibrium state. Under this assumption, \citet{yates2017msr} obtained an expression for the relative proportion of agents in each stage which can be used to closed the dynamic equations for the total density. However, since we aim to study an invasion, in which case the assumption of equilibrium does not hold, we study  the full system \eqref{eq:PDE}.}. 

\begin{figure}[h!!]
\begin{center}
	\subfigure[][]{\includegraphics[width=0.52 \columnwidth]{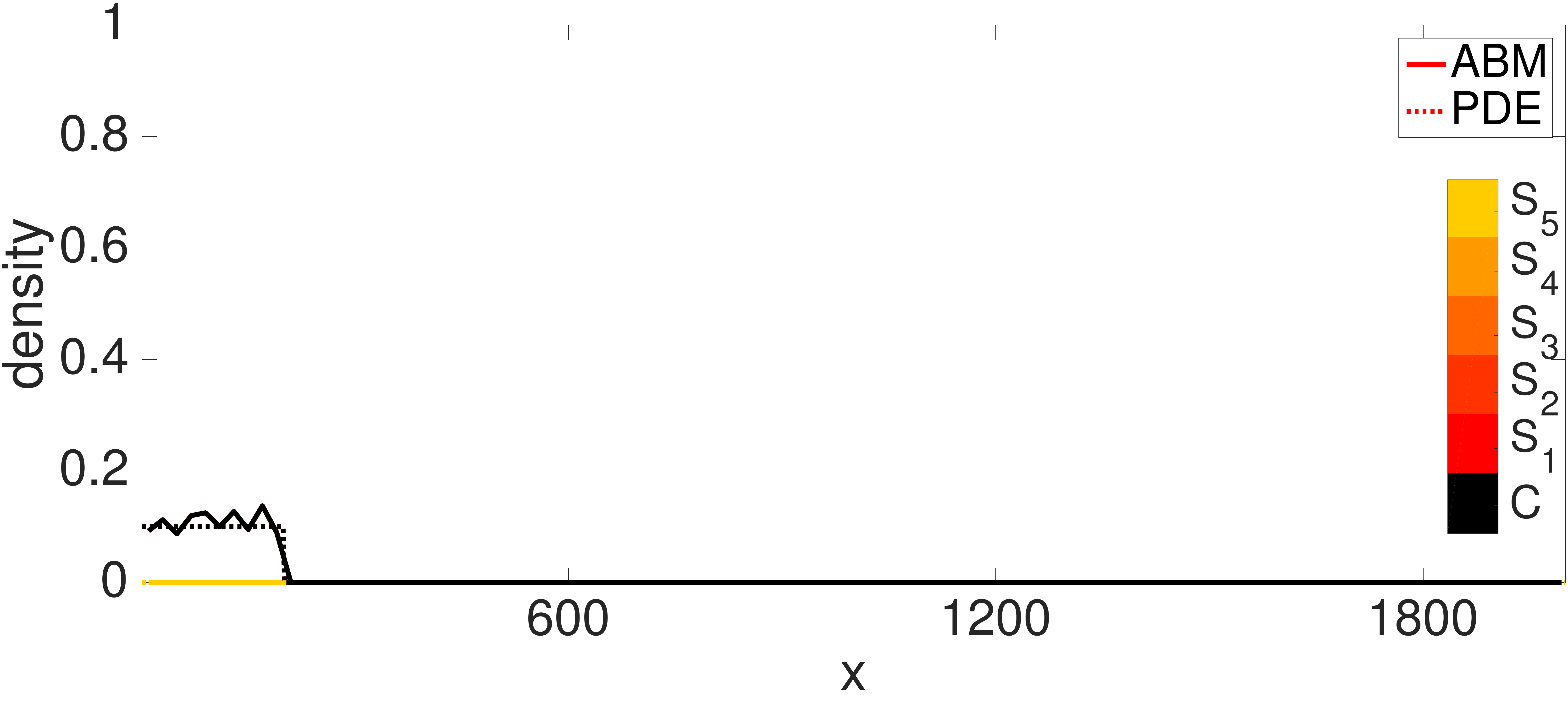} } \\
\subfigure[][]{\includegraphics[width=0.52 \columnwidth]{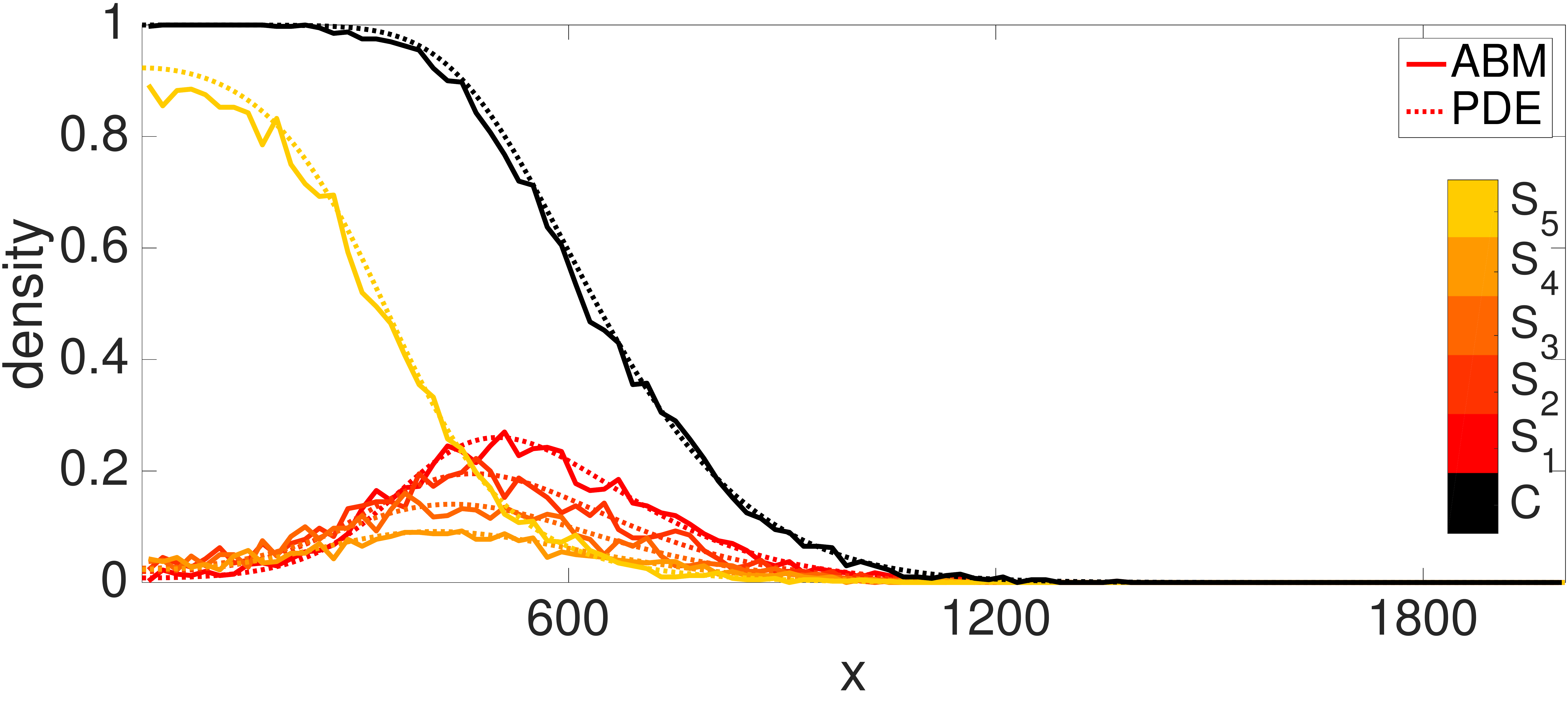} }\\
\subfigure[][]{\includegraphics[width=0.52 \columnwidth]{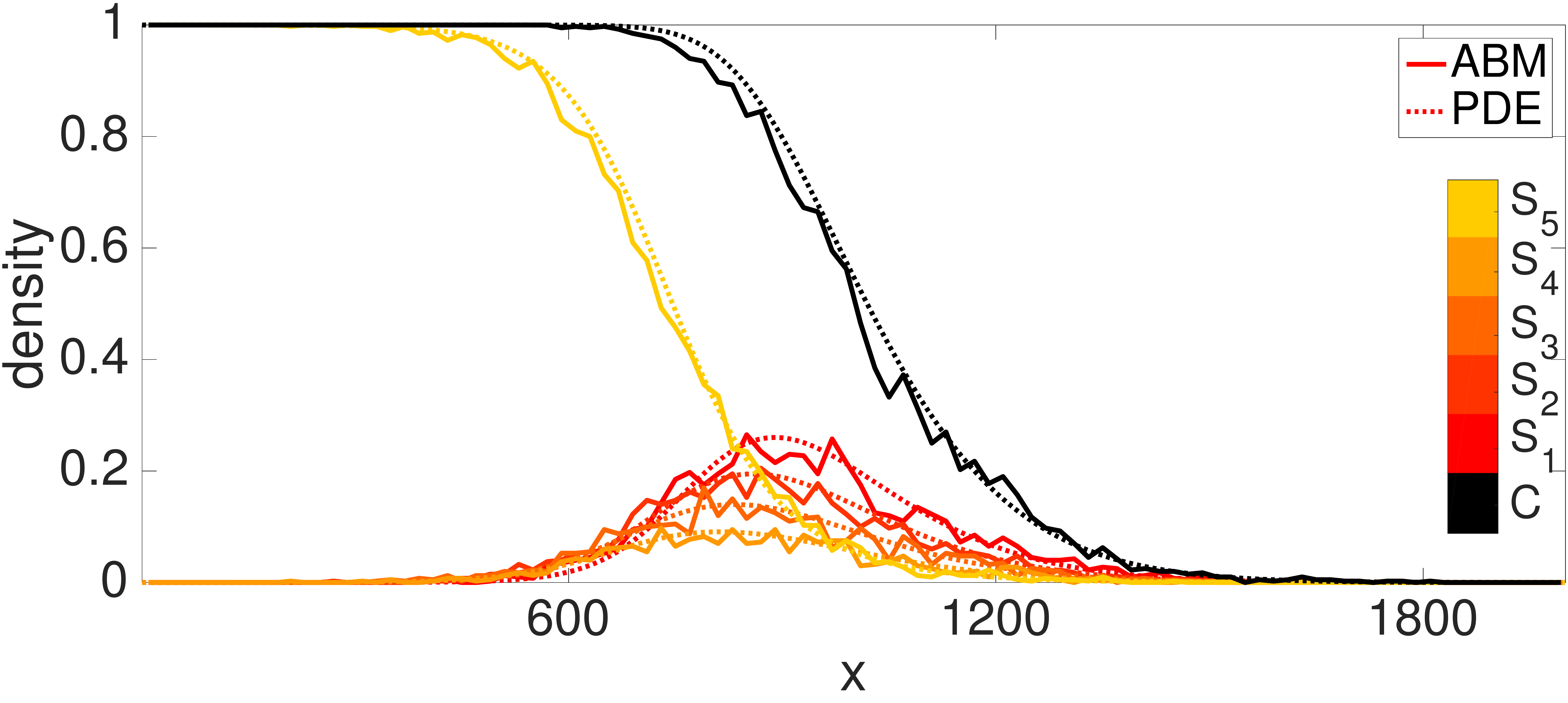}}
\end{center}
\caption{Comparison between the average column density for the ABM (full lines) and the PDE model (dotted lines) with a five-stage MSM. The panels show three snapshots of the evolution of the two models at time 0 (a), 120 (b) and 170 (c). In all cases, the profiles for the five different subpopulations are shown in different gradations of orange and the total density is plotted in black. The ABM profiles are obtained by averaging $20$ identically prepared simulations on a $2000\times 400 $ lattice. The other parameters of the models are $\Delta=20$, $\alpha=4$, $\lambda_1 =0.15$, $\lambda_2 =0.19$, $\lambda_3 =0.25$, $\lambda_4 =0.37$ and $\lambda_5 =0.75$.}
\label{fig:comparison} 
\end{figure}

We conclude this section by showing a comparison of the two models in  Figure \ref{fig:comparison}. In the example, we consider an ABM with five stages with increasing rates (the choice of the parameters is made to facilitate the visualisation of the different density profiles). In the three plots three successive snapshots are shown and the formation of the travelling wave appears clearly. As previously observed by \citet{vittadello2018mmc}, due to the presence of volume exclusion, the travelling wave solutions of the $N$ subpopulations of cells are of two qualitatively different types. The density profile of the first $N-1$ subpopulations have the form of moving pulses located at the front of the total wave with the amplitude which depends on the rate of the corresponding stage. The profile of the last stage subpopulation, instead, appears as a moving wavefront which dominates the density at the back of the total wave. 

The numerical solutions of the PDEs agree well with the average behaviour of the ABM. Therefore, we focus our attention on the the speed of the PDE model which we can investigate using an analytical approach (see Section \ref{sec:analysis}). It is important to note that the quantitative validity of our results on the PDE mode will extend to the ABM only for the range of parameters which preserves the good agreement between the two models. For example, when the rate of proliferation is large compared to the motility rate, the mean-field approximation looses accuracy and, consequentially, the speeds of the two models may differs. This is a well known phenomenon which is caused by the presence of strong the spatial correlations between occupied sites, induced by the proliferation. It is possible to derive more accurate descriptions in those case, see for example \citet{baker2010cmf, markham2013isc}, but this is beyond the scope of this paper.

\subsection{Wavespeed Analysis}
\label{sec:analysis}

System \eqref{eq:PDE} is a generalisation of the famous Fisher-KPP equation \citep{fisher1937waa}. Precisely, we can recover a Fisher-KPP equation by considering the model with a single stage, $N=1$, which is equivalent of modelling the CCTD as a single exponentially distributed random variable with rate $\lambda$. It is well know that Fisher equation admits travelling wave solutions and that the speed of invasion is given by $\nu= 2 \sqrt{\lambda D}$ \citep{fisher1937waa}. In this Section we aim to extend this result to the model of the system \eqref{eq:PDE} using the front propagation method of \citet{vansaarlos2003fpu}.\\

The system of equations \eqref{eq:PDE} has two equilibria, an unstable empty state, $S_i(x,t)\equiv 0 $ for $i=1,\dots , N$, and a stable occupied state, $S_i(x,t)\equiv 0 $ for $i=1,\dots , N-1$ and $S_N(x,t)\equiv 1$. Firstly we linearise the system about the unstable steady state, giving
\begin{equation}
\label{eq:PDE_lin}
	\begin{cases}
		\D{S_1}{t}&= D \DD{S_1}{x}+2\lambda_N  S_N - \lambda_1 S_1 \\[4 pt]
		\D{S_i}{t}&= D \DD{S_i}{x}+\lambda_{i-1} S_{i-1} - \lambda_i S_i  \qquad \quad  \text{for $i=2, \dots, N$ .}
	\end{cases}
\end{equation}
We substitute 
\begin{equation*}
	\label{eq:ansatz}
	S_i(x,t)\propto \exp \LR{- \iota  \omega(k) t+ \iota kx} \, ,
\end{equation*}
into equations \eqref{eq:PDE_lin}, where $\iota$ is the immaginary unit, $\omega(k)$ is the dispersion angular frequency of the Fourier modes and $k$ is the spatial wavenumber. Upon simplification, we obtain 
\begin{equation*}
\label{eq:modes_sub}
	\begin{cases}
		-\iota  \omega(k) &= -D k^2+2\lambda_N - \lambda_1 \\[4 pt]
		-\iota  \omega(k) &= -D k^2+\lambda_{i-1} - \lambda_i \qquad \quad  \text{for $i=2, \dots, N$.}\\[4 pt]
	\end{cases}
\end{equation*}

Following the front propagation method \citep{vansaarlos2003fpu}, the expression of the wave speed, $c$, is given by \begin{equation}
\label{eq:speed_D}
c=\frac{\IM{\omega(k^*)}}{\IM{k^*}}\, ,
\end{equation}
where $k^*=\iota  q$, with $q$ real, and such that
\begin{equation}
\label{eq:deriv_1}
	\Dstraight{\omega}{k}(k^*)=\frac{\IM{\omega(k^*)}}{\IM{k^*}} \, .
\end{equation}
Notice that we can write down $\iota  \omega(k)$ in the form
\begin{equation}
\label{eq:omega_1}
 \iota \omega(k)=k^2 D- \rho \, ,
\end{equation}
where $x$ is an eigenvalue of the matrix
\begin{equation}
\label{eq:general_matrix}
	\Lambda =\begin{bmatrix}
		-\lambda_1 && 0 && \dots && 0 && 2 \lambda_N \\
		\lambda_1 && -\lambda_2 && 0 && \dots &&  0 \\
		0 && \lambda_2 &&  -\lambda_3 && \dots && 0 \\[10pt]
		\vdots && && \ddots && \ddots && \vdots \\[10pt]
		0 && \dots && && \lambda_{N-1} && -\lambda_N
	\end{bmatrix}  \, .
\end{equation}

From expression \eqref{eq:omega_1} it follows that
\begin{subequations}
\label{eq:deriv_2}
\begin{align}
	\Dstraight{\omega}{k}(k^*)&=2 q D\, , \\
	\frac{\IM{\omega(k^*)}}{\IM{k^*}}&= \frac{q^2 D +\RE{\rho}}{q} \, .
\end{align}
\end{subequations}
By pluggin equations \eqref{eq:deriv_2} into equation \eqref{eq:deriv_1}, we obtain $q^2=\RE{\rho}/D$. Hence, from equation \eqref{eq:speed_D}, the wave speed of the invasion is given by
\begin{equation}
\label{eq:speed}
c=2 \sqrt{D\rho }\, ,
\end{equation}
where $\rho$ is the maximum real eigenvalue of $\Lambda$, defined in terms of the characteristic polynomial of the matrix $\Lambda$, $\mathcal{P}_\Lambda (x)$, as follows  
\begin{equation}
	\label{eq:rho_def}
	\rho(\Lambda)=\max \LRb{x\in \R \st \mathcal{P}_\Lambda(x)=0} \, .
\end{equation}
This shows that the problem of finding the speed of invasion of the PDE model is equivalent to computing the maximum eigenvalue $\rho(\Lambda)$ of the matrix $\Lambda$. 

\subsection{Results}
\label{sec:results}
The characteristic polynomial of the matrix $\Lambda$ can be computed directly from the matrix and it reads
\begin{equation}
	\label{eq:char_poly}
	\mathcal{P}_{\Lambda}(x)=\prod_{i=1}^N \LR{\lambda_i +x}-2 \prod_{i=1}^N \lambda_i \, .
\end{equation}
In general, an analytical formula of the roots of the polynomial function $\mathcal{P}_{\Lambda}(x)$ is not available. In this section we first consider the case of $\lambda_i=\lambda$ for $i=1, \dots, N$ for which the maximum eigenvalue $\rho(\Lambda)$ can be computed analytically. This corresponds to a special case of the general hypoexponential distribution, known as the Erlang distribution. We conclude by proving a theorem in which we state the range of speed variability for the general hypoexponential CCTD.

\paragraph{The Erlang distribution}
Consider the case $\lambda_i=\lambda$ for $i=1, \dots , N$, which corresponds the Erlang CCTD. Under this assumption, we can write down the characteristic equation of the matrix $\Lambda$, using formula \eqref{eq:char_poly}, as
\begin{equation}
	\label{eq:char_poly2}
	\LR{\lambda +x}^N=2  \lambda^n \, .
\end{equation}
The eigenvalues of $\Lambda$ are then given by the solutions of equation \eqref{eq:char_poly2} which are $
	x_j= \lambda \LR{\xi^j \sqrt[N]{2} -1} $ for $j=1, \dots , N$, where $\xi=\exp \LR{2 \pi \iota  /N}$ is the $N$-th root of unity.  Hence, we obtain that  
\begin{equation}
\label{eq:xstar_erlang}
	\rho(\Lambda)=\lambda \LR{\sqrt[N]{2} -1} \, .
\end{equation}
By substituting the expression \eqref{eq:xstar_erlang} into equation \eqref{eq:speed} we obtain the formula for the speed of invasion for the model with Erlang distribution
\begin{equation}
	\label{eq:speed_erlang}
	c=2 \sqrt{D\lambda \LR{\sqrt[N]{2}-1}} \, . 
\end{equation}
Notice that for $N=1$, which corresponds to exponential CCTD, we recover  the well known expression of the speed for the Fisher-KPP equation.

\paragraph{The general hypoexponential distribution}
For the case of a general hypoexponential distribution, there is no analytical formula for the expression of the maximum real eigenvalue of the matrix $\Lambda$. However, we find that the Erlang case and the exponential case, for which we do have the analytical formula of the speed, correspond to the lower and upper bound (respectively) for the speed of travelling waves with hypoexponential CCTD and a given total proliferation rate, $\bar{\lambda}$. This result follows directly from the following theorem on the range of $\rho(\Lambda)$.

\paragraph{Theorem 2}
Let $\rho(\Lambda)$ be defined by equation \eqref{eq:rho_def} as the maximum real eigenvalue of the matrix $\Lambda$. Then
\begin{equation} 	
\label{eq:theorem}
\bar{\lambda} N\LR{\sqrt[N]{2}-1} \le \rho(\Lambda) < \bar{\lambda} \, ,
\end{equation}
where $\bar{\lambda}=\LR{\sum_{i=1}^N 1/\lambda_i}^{-1}$. 

\begin{proof}
Let $\mu_i=1/\lambda_i$ for every $i=1, \dots, N$. By writing the characteristic equation $\mathcal{P}_{\Lambda}(x)=0$ in terms of the parameters $\mu_i$ and upon rearranging, we obtain
\begin{equation}
\label{eq:char_eq_mu}
	\prod_{i=1}^N\LR{\mu_i x +1}=2 \, .
\end{equation} 
We can write $\rho(\Lambda)=\rho(\mu_1, \dots , \mu _N)=\maxeig$ as
\begin{equation}
	\label{eq:rho_def_proof}
	\maxeig=\max\LRb{x\in \R \st \prod_{i=1}^N\LR{\mu_i x +1}=2} \, ,
\end{equation}
for every $\underbar{$\mu$}\in \LRb{\R_>}^N$. It is easy to observe that $\maxeig$ is a positive continuous function and we can extend the definition \eqref{eq:rho_def_proof} to $\underbar{$\mu$}\in \LRb{\R_\ge}^N \,\setminus \LRb{(0, \dots, 0)}$, by continuity.

Now fix $\bar{\lambda}=\LR{\sum_{i=1}^N \mu_i}^{-1}$; without loss of generality we can take $\sum_{i=1}^N\mu_i =1$, whence \eqref{eq:theorem} becomes $N\LR{\sqrt[N]{2}-1}\le \maxeig <1$.  The case of general $\bar{\lambda}$ follows by multiplying by rescaling factor. Since $\rho$ is a continuous function, we aim to find the stationary points of $\maxeig$ in the $N$-dimensional simplex:
\begin{equation}
	\label{eq:simplex}
	\mathcal{U}_N=\LRb{\LR{\mu_1, \dots , \mu_N} \in (0,1]^N \, \Big| \, \sum_{i=1}^N \mu_i =1 } \, .
\end{equation}
We apply the Lagrange multipliers method. Hence we study the Lagrangian function given by
\begin{equation}
	\label{eq:lagrangian}
	\lagra(\mu_1, \dots , \mu_N , \sigma)= \maxeig + \sigma \LR{\sum_{i=1}^N \mu_i -1} \, .
\end{equation}
Throughout we adopt the notation $\lagra_j=\D{\lagra}{\mu_j}$ and $\rho_j=\D{\rho}{\mu_j}$. By imposing $\lagra_j=0$ we obtain
\begin{equation}
\label{eq:lagr_condition}
	\rho_j=-\sigma \, ,
\end{equation}
for all $j=1, \dots , N$. We can now differentiate equation \eqref{eq:char_eq_mu} respect to $\mu_j$, which gives us
\begin{equation}
\label{eq:pol_char_derived}
	0=\sum_{i=1}^N \prod_{k \ne i}\LR{1+\mu_k \rho} \LR{\rho\, \delta_{i,j} +\mu_i \rho_j} 
\end{equation}
	where $\delta_{i,j}$ denotes the Kronecker delta. If we multiply and divide each term of the right-hand side of equation \eqref{eq:pol_char_derived} by $(1+\mu_i\rho)$, we obtain
\begin{align}
\label{eq:proof_step}
		0&=\sum_{i=1}^N \frac{\rho\, \delta_{i,j} +\mu_i \rho_j }{1+\mu_i \rho}\nonumber  \\ 
	&= \frac{\rho}{1+\mu_j \rho}+\rho_j \sum_{i=1}^N \frac{\mu_i  }{1+\mu_i \rho} \, ,
\end{align}
By combining equations \eqref{eq:lagr_condition} and \eqref{eq:proof_step} we gain a condition on the coordinate $\mu_j$ of the stationary points, namely  
\begin{equation}
\label{eq:stationary_condition}
	\frac{\rho}{1+\mu_j \rho}=\sigma\sum_{i=1}^N \frac{\mu_i  }{1+\mu_i \rho} \, .
\end{equation}
Notice that equation \eqref{eq:stationary_condition} holds for every $j=1, \dots , N$ and it is independent of $j$, hence the only stationary point of $\maxeig$ in the simplex $\mathcal{U}_N$ is the given by the centre $\underbar{$\mu$}^*_N=\LR{1/N, \dots , 1/N}$.  

To conclude we need study the value of $\maxeig$ on the boundary of the simplex, defined as
\begin{equation}
\partial \mathcal{U}_N= \LRb{\LR{\mu_1, \dots , \mu_N} \in \LRs{0,1}^N \, \Big| \, \sum_{i=1}^N \mu_i =1 \,  \, \text{and $\mu_j=0$, $\exists j\in \LRb{1, \dots , N}$}  }\, .
\end{equation}
Let's consider the elements of $\partial \mathcal{U}_N$ with exactly $n$ non-zero coordinates, with $n=1,\dots, N-1$. Without loss of generality we can focus on the points of the form 
\begin{equation}
\label{eq:induction_form}
\LR{\mu_1, \dots, \mu_{n},0,\dots, 0} \in \partial \mathcal{U}_N \, ,
	\end{equation}
where $\LR{\mu_1, \dots, \mu_{n}} \in \mathcal{U}_{n}$. Notice that the $\maxeig$ is well defined in such points by continuity, as observed before. By repeating the Lagrange multiplier method in the sub-simplex $\mathcal{U}_n$, we find that the only stationary point of $\maxeig$ of the form \eqref{eq:induction_form} is the one with $\mu_1=\mu_2=\cdots =\mu_n$, i.e.:  
\begin{equation}
\label{eq:stat_form_proof}	
\underbar{$\mu$}^*_n =(\underbrace{1/n,\dots, 1/n }_n, 0, \dots, 0) \in \partial \mathcal{U}_N\, . 
\end{equation}
This holds for every $n=1,\dots, N-1$, so we can write all the stationary points of $\maxeig$ in $\partial \mathcal{U}_N$ upon permutation of the coordinates in the form \eqref{eq:stat_form_proof}. 

All the stationary points $\underbar{$\mu$}^*_n$, for $n=1, \dots N$, correspond to an Erlang distribution for which we can compute the expression of $\rho$  directly from the definition \eqref{eq:rho_def_proof} as 
\begin{equation}
\label{eq:value_stat_proof}
\rho(\underbar{$\mu$}^*_n) = n\LR{\sqrt[n]{2}-1} \, ,
\end{equation}
for $n=1, \dots, N$. The right-hand side of equation \eqref{eq:value_stat_proof} is a decreasing function of $n$. We deduce that the centre of the simplex, $\mu^*_N\in \mathcal{U}_N$, corresponds to the global minimum, i.e. for all $\underbar{$\mu$}\in \mathcal{U}_N$ 
\begin{equation}
\maxeig\ge\rho(\underbar{$\mu$}^*_N)=N\LR{\sqrt[N]{2}-1}\, .	
\end{equation}
Finally, $\mu^*_1 \in \partial \mathcal{U}_N$ and all the points obtained by permuting its coordinates, correspond to supremum points, i.e. for all $\underbar{$\mu$}\in \mathcal{U}_N$ 
\begin{equation} 
\maxeig <\rho(\underbar{$\mu$}^*_1)=1 \, .
\end{equation}

\end{proof}

It is immediate to interpret the result of the Theorem 2 in terms of invasion speeds. In particular, by using equation \eqref{eq:speed},  together with the two inequalities \eqref{eq:theorem}, we deduce that the speed of the invasion of the PDE model with diffusion coefficient $D$ and a general $N$-stage representation of the CCTD with total growth rate given by $\bar{\lambda}$, lies in the interval 
\begin{equation}
\label{eq:range_N}
c \in \left[2 \sqrt{D \bar{\lambda} N\LR{\sqrt[N]{2}-1}} , \,2 \sqrt{D \bar{\lambda}}\right) \, .
\end{equation}

We can generalise this result even further by taking the limit as $N\rightarrow \infty$ in the right-hand side of equation \eqref{eq:range_N}. Hence we obtain a general interval which holds for any multi-stage representation, regardless of the number of stages, which reads 
\begin{equation}
\label{eq:range_limit}
	c \in \left(2 \sqrt{D \bar{\lambda} \ln 2} , \,2 \sqrt{D \bar{\lambda}}\right) \, ,
\end{equation}
where we used $N\LR{\sqrt[N]{2}-1}=\ln 2 + \mathcal{O}\LR{N^{-1}}$.

\begin{figure}[h!!]
\begin{center}
	\includegraphics[width=0.65 \columnwidth]{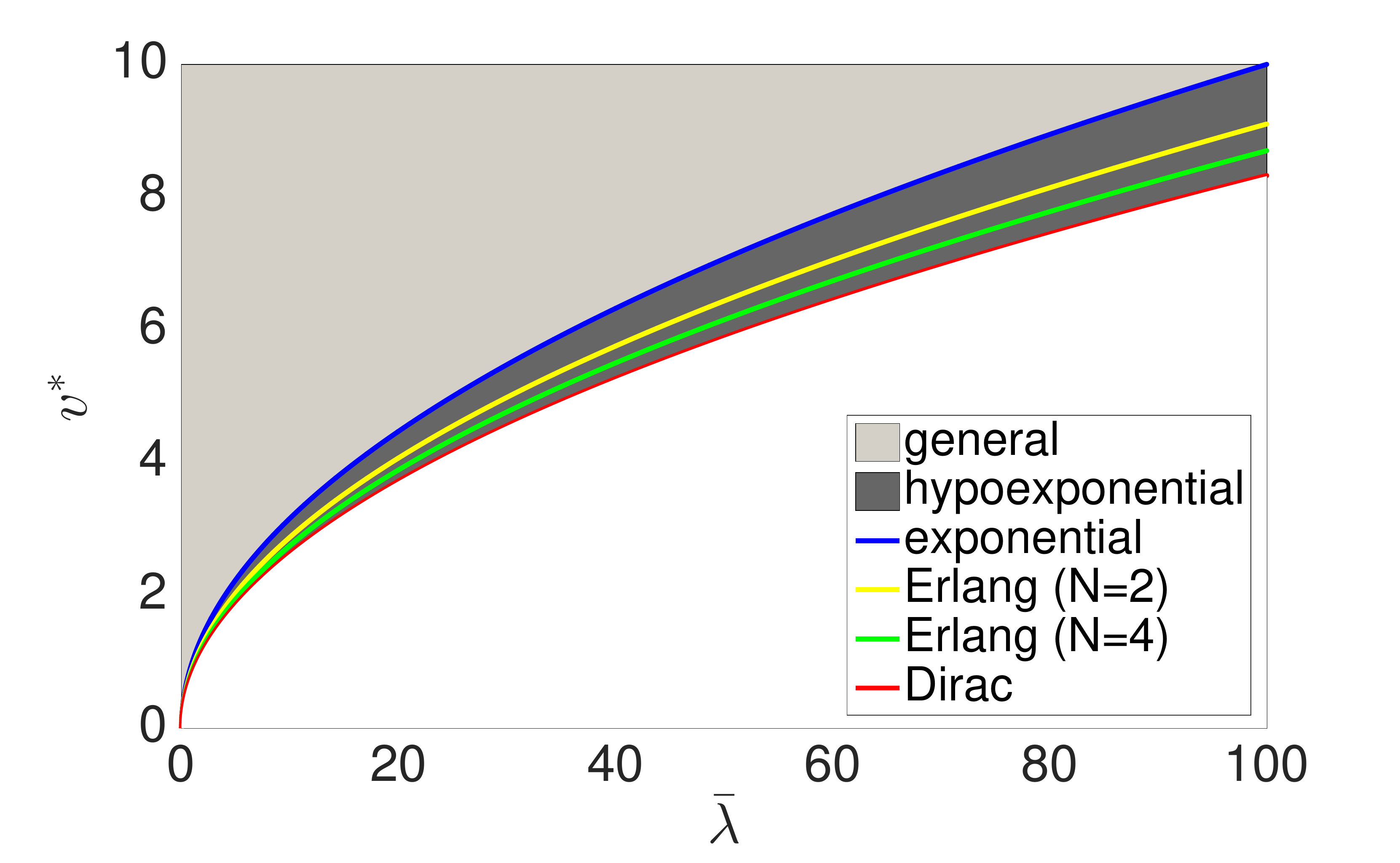} 
\end{center}
\caption{Illustration of the range of invasion speeds for a fixed mean proliferation rate and diffusion coefficient, $D=1$. The two coloured regions represent the range of speed for a general CCTD. The dark grey subregion highlights the range of speeds for hypoexponential CCTDs. The global minimum speed is obtained by using the Dirac distribution (red line). The exponential CCTD (blue line) is the hypoexponential distribution which leads to maximum speed. There is no upper bound for the general case. Two examples of Erlang CCTDs with two stages (yellow line) and four stages (green line) are also shown.}
\label{fig:illustration} 
\end{figure}

Notice that the lower bound of the interval \eqref{eq:range_limit} is equivalent to the lower bound for the general CCTD, obtained in \eqref{eq:range_general} of Section \ref{sec:age_struc}. This can be intuitively understood by observing that, as we let number of stages of an hypoexponential distribution go to infinity while keeping the total rate, $\bar{\lambda}$, fixed, the variance of the distribution tends to zero. Consequently, the distribution converges to a Dirac function concentrated in the mean, $\bar{\mu}=\bar{\lambda}^{-1}$, which we have proved in Section \ref{sec:age_struc} to be the distribution corresponding to the minimum invasion speed. In Figure \ref{fig:illustration} we summarise our findings about the range of invasion speed for different CCTD through a graphical representation.

 \section{Conclusion}
\label{sec:conclusions}

In this work we investigated the quantitative effect of implementing a realistic CCTD into models of cell invasion. Firstly, we derived a general result from a generalised version of the Fisher-KPP equation. Then we investigated the case of MSMs by implementing a simple ABM of cells undergoing undirected migration and proliferation by division, in which the time between successive divisions is modelled using a multi-stage representation (i.e. the CCTD is hypoexponential). By studying a continuous version of the ABM, we connected the type of CCTD to the speed of the corresponding invasion. 

The results indicate that, for a fixed mean division time, the minimum speed of invasion is obtained by the Dirac distribution, while there is no upper bound. In other words, the invasion can be, in general, infinitely fast. However, when we focus our attention to the case of MSMs, which are known to represent well the experimental CCTD, our analysis shows that the speed can vary in a bounded interval (see Figure \ref{fig:illustration}). More precisely, we show that the maximum invasion speed is reached by adopting an exponential CCTD, which leads to the classic the Fisher-KPP model. On the other hand, the minimum speed is obtained by partitioning the CCTD into multiple exponential stages with identical rates, which corresponds to the case of Erlang CCTD. Finally, by considering the limiting case of infinitely many stages, we find that the infimum value of the speed for the class of hypoexponential CCTD coincides with the global minimum for a general CCTD.

The results indicate the invasion speed changes with the variance of the CCTD, i.e. decreasing the variance in the proliferation time distribution leads to slower invasion. We found that the maximum reduction in comparison to the classical formula for the Fisher-KPP model, is given by a multiplicative factor of $\sqrt{\ln 2}\approx 0.83$. Whilst interpreting this result in the context of experimental data is beyond the aim of this work, we want to stress that for number of stages $N\gg 1$, which is typically the case for experimentally observed distributions \citep{golubev2016aie, yates2017msr,chao2018ecc}, the speed converges to the lower bound of equation \eqref{eq:range_limit} with order given by $\mathcal{O}\LR{N^{-1}}$. This suggests that, with the only information of the mean  of the CCTD (equivalently, the total rate), including the factor $\sqrt{\ln 2}$ in the formula for the speed leads to a more accurate estimation than the classic expression of Fisher-KPP.

An important question that remains unanswered is the role of motility heterogeneity within the cell cycle. Many experimental studies have found that the motility of a cell can depend on its cell cycle phase \citep{vittadello2018mmc}. For example, during the mitotic phase, cells tend to reduce their movement \citep{mort2016rdm}. In order to investigate this phenomenon in the light of the invasion speed, we could modify our model to allow different diffusion coefficients, $D_i$ for $i=1, \dots, N$, for each stage in the system \eqref{eq:PDE}. Another aspect of the cell movement that can vary within the cell cycle is the directional persistence. Our models do not incorporate directional persistence of cells. However, it is possible to combine a MSM with existing models of directional persistence \citep{codling2008rwm, gavagnin2018mpm}. Unfortunately, the application of the front propagation method of \citet{vansaarlos2003fpu} (see Section \ref{sec:analysis}) to these models leads to a dead end and it may be necessary to study the problem using a different approach.  We will investigate this in future research.

%
%

\bibliography{database}{}
\bibliographystyle{unsrtnat}

\end{document}